\documentclass[fleqn,usenatbib]{mnras}

\usepackage{newtxtext,newtxmath}

\usepackage[T1]{fontenc}

\DeclareRobustCommand{\VAN}[3]{#2}
\let\VANthebibliography\thebibliography
\def\thebibliography{\DeclareRobustCommand{\VAN}[3]{##3}\VANthebibliography}

\usepackage{graphicx}	
\usepackage{amsmath}	
\usepackage{amssymb}	

\DeclareGraphicsExtensions{.pdf,.png,.jpg}

\newcommand{\chandra}{\textit{Chandra}}

\newcommand{\xmm}{\textit{XMM-Newton}}
\newcommand{\xspec}{{\sc xspec}}

\newcommand{\apec}{{\it apec}}
\newcommand{\bvapec}{{\it bvapec}}
\newcommand{\windtabs}{{\it windtabs}}
\newcommand{\vwindtab}{{\it vwindtab}}
\newcommand{\tbabs}{{\it tbabs}}

\newcommand{\zpup}{$\zeta$~Pup}
\newcommand{\xper}{$\xi$~Per}
\newcommand{\zori}{$\zeta$~Ori}
\newcommand{\eori}{$\epsilon$~Ori}
\newcommand{\zoph}{$\zeta$~Oph}
\newcommand{\sgr}{$9$~Sgr}

\newcommand{\Ro}{\ensuremath{{R_{\mathrm o}}}}

\newcommand{\Rstar}{\ensuremath{{R_{\ast}}}}

\newcommand{\Rsun}{\ensuremath{\mathrm {R_{\sun}}}}

\newcommand{\Msunyr}{\ensuremath{{\mathrm {M_{\sun}~{\mathrm yr^{-1}}}}}}
\newcommand{\ciao}{CIAO}

\newcommand{\rgs}{RGS}
\newcommand{\meg}{MEG}
\newcommand{\heg}{HEG}

\newcommand{\vinf}{\ensuremath{v_{\infty}}}

\newcommand{\Lx}{\ensuremath{L_{\rm X}}}
\newcommand{\Lbol}{\ensuremath{L_{\rm bol}}}

\newcommand{\Mdot}{\ensuremath{\rm \dot{M}}}

\newcommand{\Sigmastar}{${\Sigma}_{\ast}$}

\newcommand{\beq}{\begin{equation}}
\newcommand{\eeq}{\end{equation}}
\newcommand{\beqa}{\begin{eqnarray}}
\newcommand{\eeqa}{\end{eqnarray}}

\title[{\it Chandra} O star wind emission and absorption]{{\it Chandra} grating spectroscopy of embedded wind shock X-ray emission from O stars shows low plasma temperatures and significant wind absorption}
\author[D.Cohen et al.]{David H.\ Cohen,$^{1}$\thanks{E-mail:
    dcohen1@swarthmore.edu} Winter Parts,$^{1}$ Graham M. Doskoch,$^{1}$  
\newauthor Jiaming Wang,$^{1}$ V\'{e}ronique Petit,$^{2}$ Maurice A. Leutenegger,$^{3}$ 
\newauthor Marc Gagn\'e$^{4}$  \\
  $^{1}$Swarthmore College, Department of Physics and Astronomy, Swarthmore, Pennsylvania 19081, USA\\
   $^{2}$University of Delaware, Department of Physics and Astronomy, Newark,
  Delaware 19716, USA \\
   $^{3}$NASA/Goddard Space Flight Center, Code 662, Greenbelt, Maryland 20771, USA \\
  $^{4}$West Chester University, Department of Earth and Space Sciences, West Chester, Pennsylvania 19383, USA \\
}

\date{}

\pubyear{2021}

\begin{document}
\label{firstpage}
\pagerange{\pageref{firstpage}--\pageref{lastpage}}
\maketitle

\begin{abstract}
We present a uniform analysis of six examples of embedded wind shock (EWS) O star X-ray sources observed at high resolution with the \chandra\/ grating spectrometers. By modeling both the hot plasma emission and the continuum absorption of the soft X-rays by the cool, partially ionized bulk of the wind we derive the temperature distribution of the shock-heated plasma and the wind mass-loss rate of each star. We find a similar temperature distribution for each star's hot wind plasma, consistent with a power-law differential emission measure, $\frac{d\log EM}{d\log T}$, with a slope a little steeper than -2, up to temperatures of only about $10^7$ K. The wind mass-loss rates, which are derived from the broadband X-ray absorption signatures in the spectra, are consistent with those found from other diagnostics. The most notable conclusion of this study is that wind absorption is a very important effect, especially at longer wavelengths. More than 90 per cent of the X-rays between 18 and 25 \AA\/ produced by shocks in the wind of \zpup\/ are absorbed, for example. It appears that the empirical trend of X-ray hardness with spectral subtype among O stars is primarily an absorption effect.
\end{abstract}


\section{Introduction} \label{sec:intro}

The dense and highly supersonic radiation-driven winds of O stars generate thermal soft X-ray emission from the cooling of shock-heated plasma. These embedded wind shocks (EWS) are thought to be produced by the line deshadowing instability (LDI) intrinsic to radiation-driven flows in which momentum transfer is mediated by spectral lines \citep{OCR1988,Feldmeier1997,Sundqvist2018}. Although this picture is widely accepted, it is far from completely tested and characterized. There are numerous physically plausible model ingredients that can affect the amount of shock-heated wind plasma and its temperature distribution. In this paper we present a comprehensive and uniform analysis of the \chandra\/ grating spectra of six O stars with EWS X-ray emission in order to characterize the wind plasma temperature distribution in ways that will be useful for constraining shock heating and cooling models.

From the observational perspective, a trend of X-ray hardness correlated with spectral subtype has been noted in the ensemble of \chandra\/ grating spectra of O and early B stars, interpreted as an underlying correlation between X-ray plasma temperature and stellar effective temperature or wind strength \citep{wnw2009}. \citet{Leutenegger2010} suggested that wind absorption could account for all or most of the trend and presented a new, simple but realistic model of X-ray transport in the EWS scenario. This built on earlier work exploring the broadband wind absorption of X-rays in OB stars \citep{Waldron1984,Hillier1993,Cohen1996,OC1999,Waldron2007}. Previous analyses of \chandra\/ grating spectra of hot stars have found a plasma temperature distribution \citep{ws2005} or a shock distribution \citep{zp2007,Cohen2014b} skewed heavily toward lower temperatures, but have not identified a clear trend with stellar or wind properties. 

We aim here to present a uniform analysis of O stars with high quality \chandra\/ grating spectra whose emission is thought to be dominated by the EWS mechanism, and not by magnetically channeled wind shocks (MCWS) or colliding wind shocks (CWS) in a binary system. This analysis enables direct comparison of emission temperature distributions of different stars and includes both realistic radiation transport through the wind -- accounting for wind ionization and the spatial distribution of the emitting and absorbing plasma -- and non-solar CNO abundances when applicable. 


\begin{figure*}
	\includegraphics[angle=0,width=0.86\textwidth]{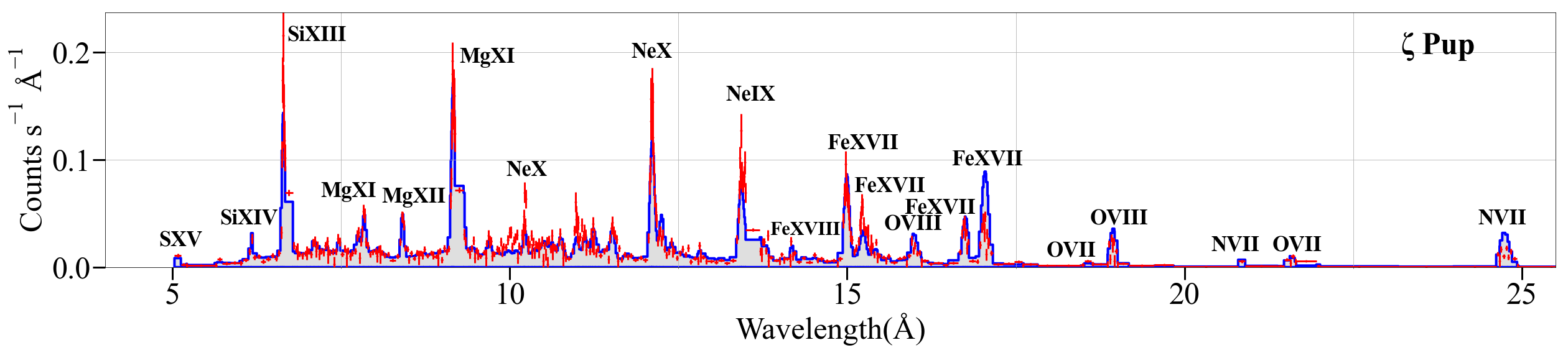}
	\includegraphics[angle=0,width=0.86\textwidth]{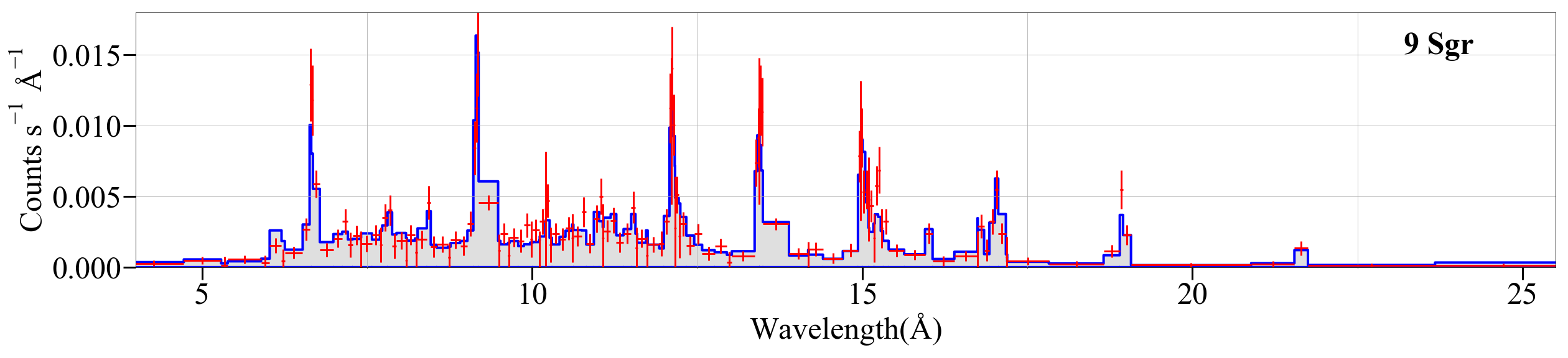}
	\includegraphics[angle=0,width=0.86\textwidth]{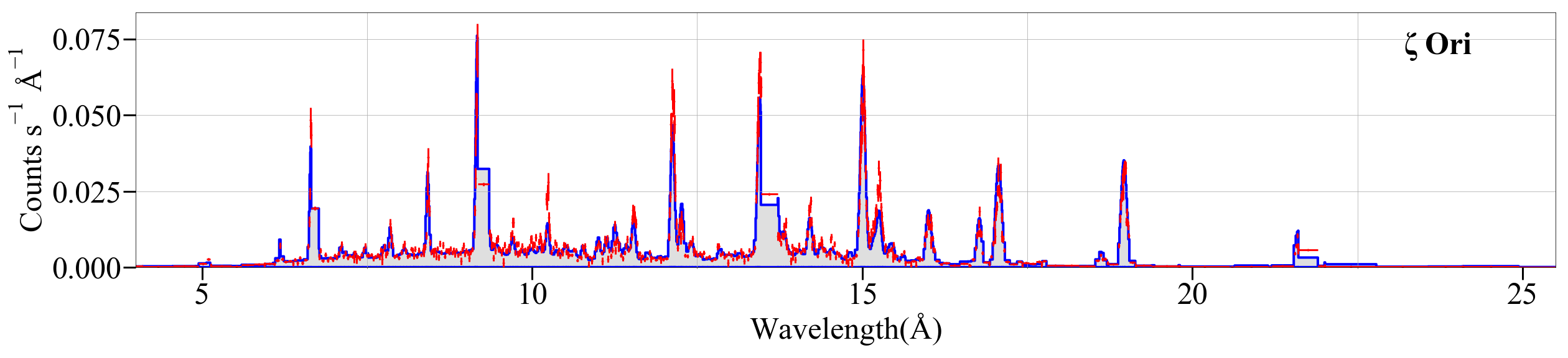}
	\includegraphics[angle=0,width=0.86\textwidth]{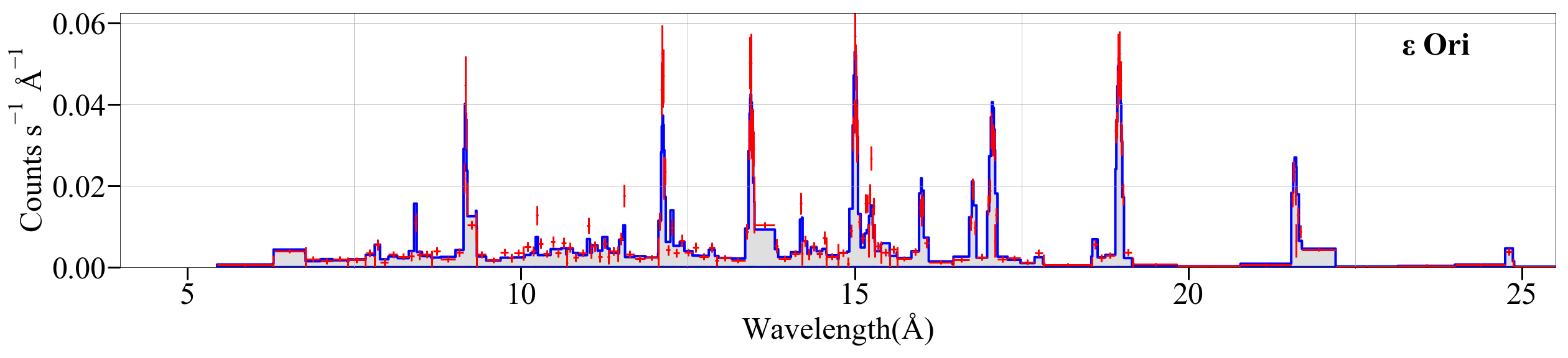}
	\includegraphics[angle=0,width=0.86\textwidth]{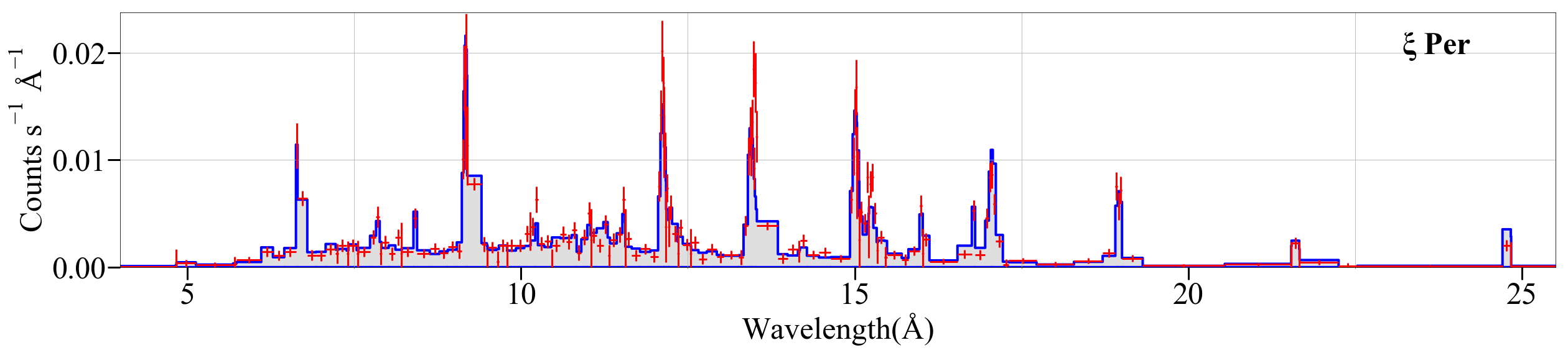}
	\includegraphics[angle=0,width=0.86\textwidth]{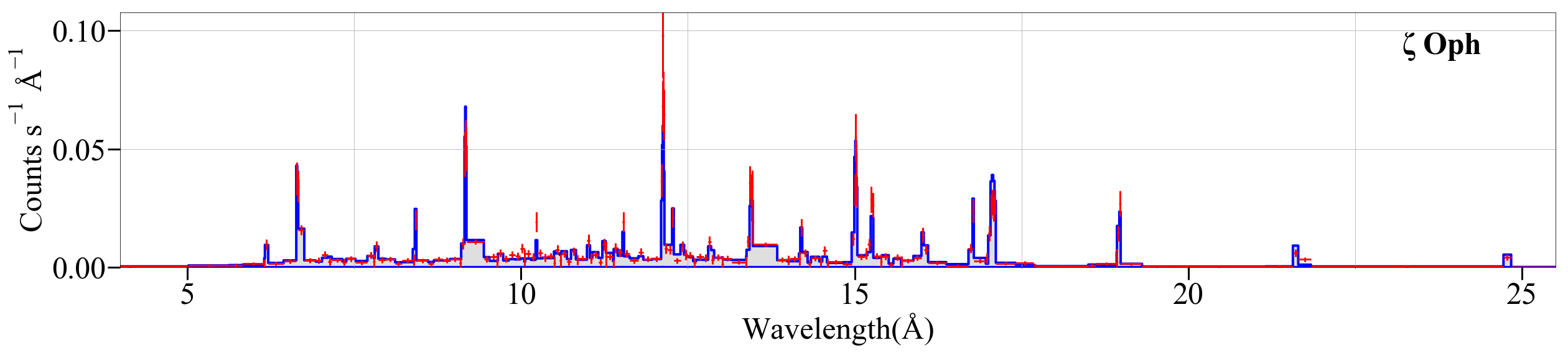}
	\caption{ The co-added and grouped MEG spectra (red, with error bars) for $\zeta$ Pup, 9 Sgr, $\zeta$ Ori, $\epsilon$ Ori, $\xi$ Per, and $\zeta$ Oph (top to bottom, ordered by theoretical wind mass-loss rate). The shaded (blue) histograms are the best-fit models presented in \S4. Noticeable emission lines are labeled according to their parent ions in the top panel. The labels apply to all the spectra.
	}   
	\label{fig:MEGdata}
\end{figure*}


\begin{figure*}
	\includegraphics[angle=0,width=0.86\textwidth]{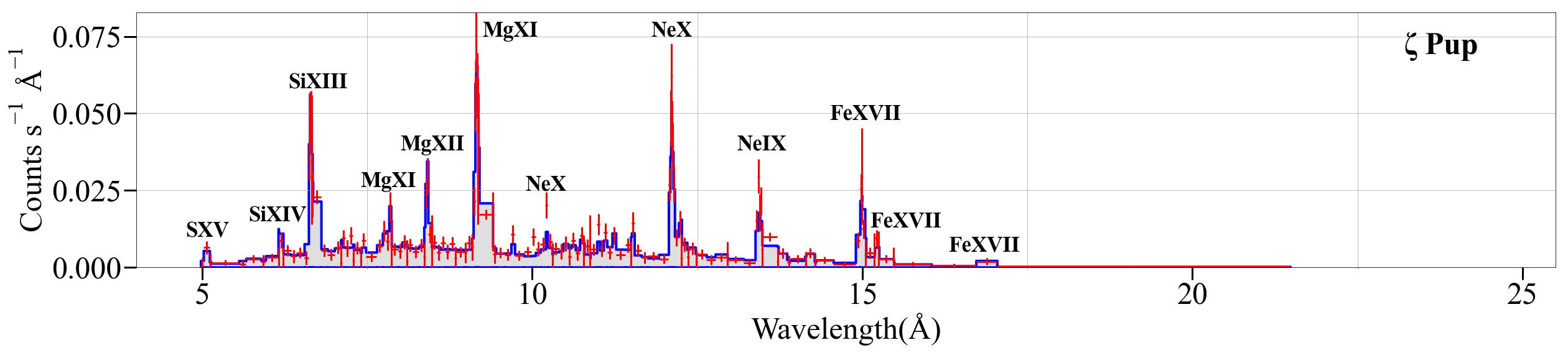}
	\includegraphics[angle=0,width=0.86\textwidth]{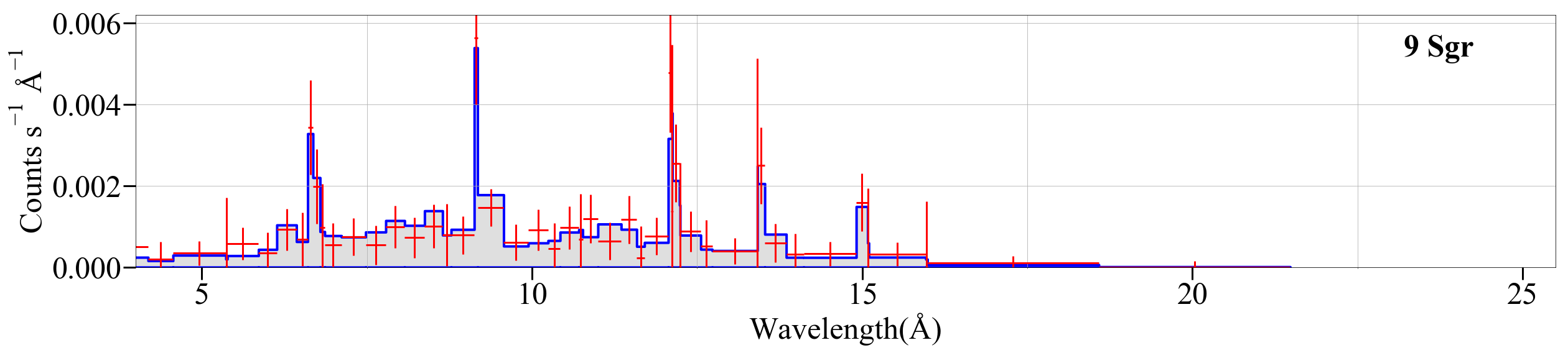}
	\includegraphics[angle=0,width=0.86\textwidth]{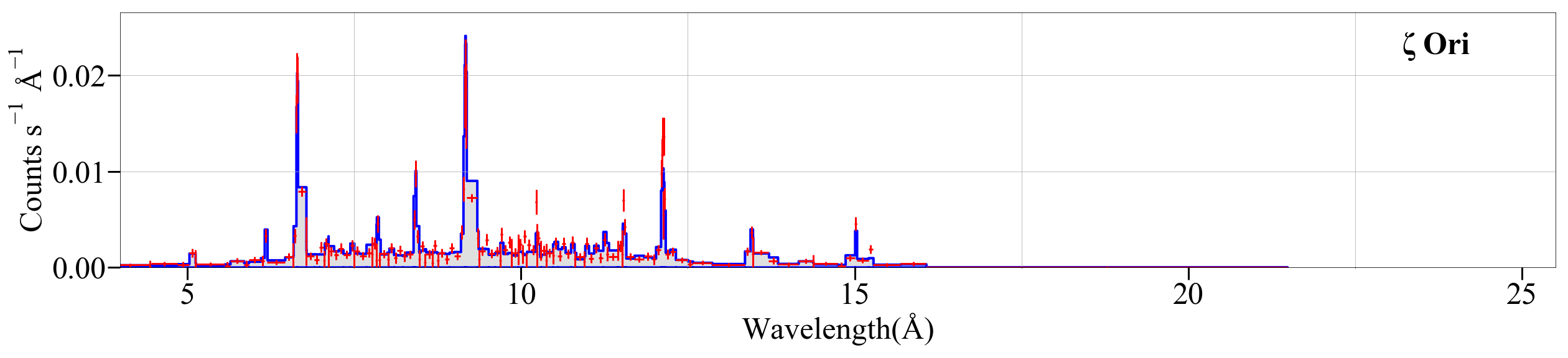}
	\includegraphics[angle=0,width=0.86\textwidth]{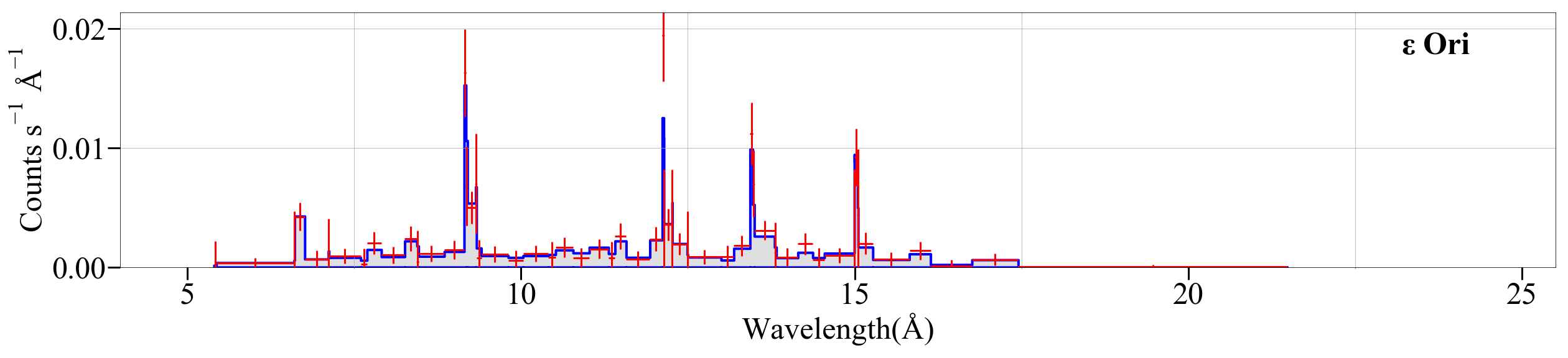}
	\includegraphics[angle=0,width=0.86\textwidth]{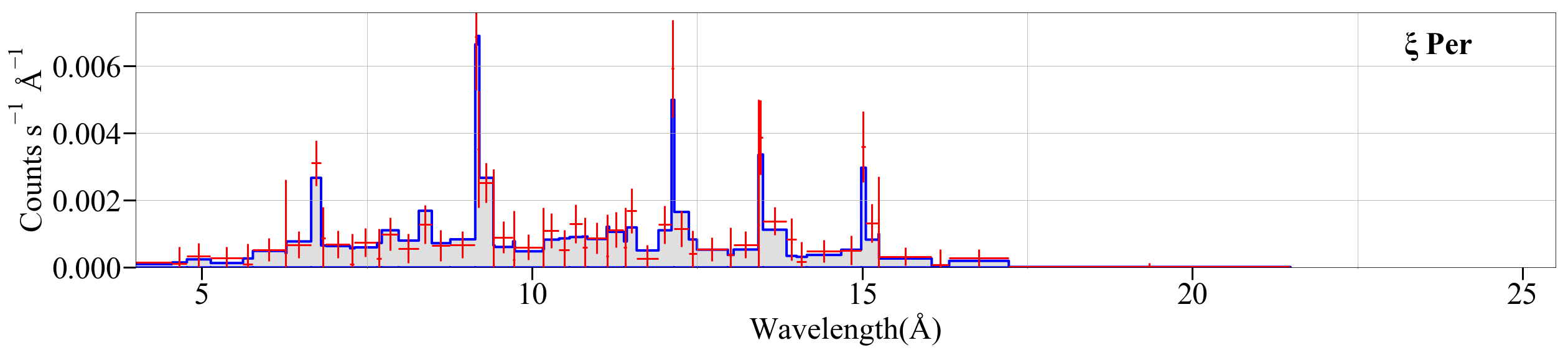}
	\includegraphics[angle=0,width=0.86\textwidth]{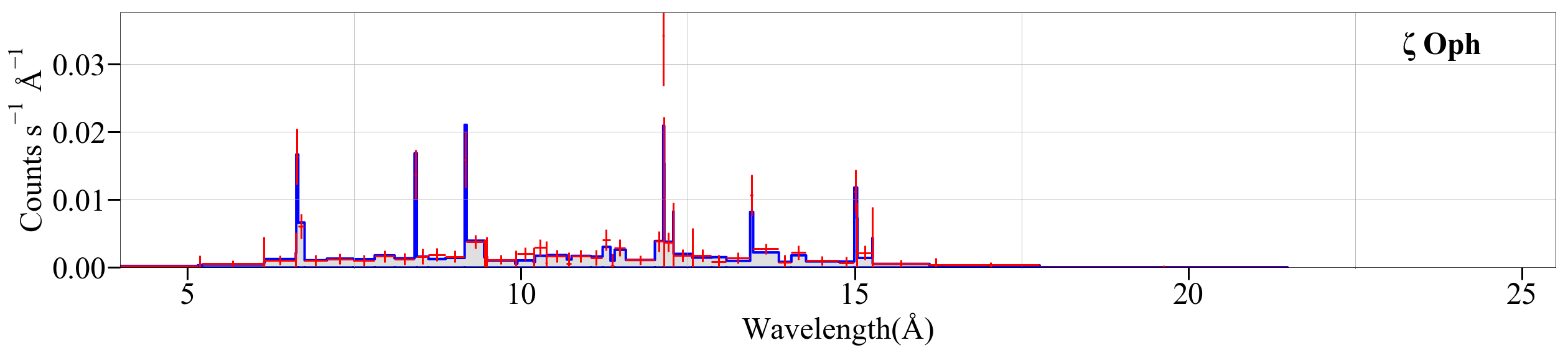}
	\caption{Same as for Fig.\ \ref{fig:MEGdata}, but showing the HEG data.
	}   
	\label{fig:HEGdata}
\end{figure*}

Wind absorption and the shocked plasma emission temperature distribution are quite degenerate when it comes to controlling the overall appearance of broadband X-ray spectra of O stars, with both higher temperatures and more absorption leading to harder spectra. The earliest O star in the Galaxy, HD 93129A (O2If) has the hardest X-ray spectrum in the \citet{wnw2009} sample, but detailed, simultaneous broadband modeling of its medium-resolution zeroth order \chandra\/ spectrum and its X-ray emission line profiles show that the plasma temperature is not in fact high but rather that there is a significant amount of spectral hardening due to wind absorption (Cohen et al. 2011). Indeed, 80 per cent of the X-rays emitted in the \chandra\/ bandpass are absorbed before they can escape the wind of HD 93129A. So wind absorption should be important in interpreting other O stars' observed X-ray spectral energy distributions as well, and it may have a systematic effect on altering derived plasma temperature distributions when compared to those derived without accounting for wind absorption. By accurately accounting for absorption, the derived emission-measure distributions as a function of temperature can be more readily compared to numerical models of embedded wind shocks.

We use the {\it windtabs} stellar wind X-ray absorption model \citep{Leutenegger2010}, which treats the radiation transport of a distributed emitter embedded in an absorbing medium and includes a user-specified opacity model for the wind. Both of these features represent a significant improvement over treating wind absorption as exponential attenuation by a neutral medium as ISM absorption models do. The {\it windtabs} model is implemented in \xspec\/ \citep{Dorman2001} and runs quickly and so can be used as easily as the standard ISM absorption models. We couple this wind absorption model to a multi-temperature, line-broadened, variable abundance, {\it bvapec} thermal spectral emission model \citep{Foster2012}. For both conceptual simplicity and to facilitate comparisons among stars we approximate the continuous temperature distribution -- the differential emission measure (DEM) -- as a sum of six fixed temperatures, the choice of which is based both on the emissivity functions of the lines measured with the \chandra\/ gratings and on empirical testing with different temperature values. Fitting this wind emission and absorption model enables us to measure the DEM and also to measure the wind mass-loss rate as well as nitrogen and oxygen abundances. 

We describe the sample and data in \S2. In \S3 we elaborate on the modeling summarized here in the Introduction. We present our results in \S4, discuss the implications in \S5, and summarize our main conclusions in \S6. 


\section{The sample and data} \label{sec:data}


\begin{table*}
  \caption{Stellar and wind properties}
\begin{tabular}{cccccccccc}
  \hline
  Star & HD \# & Spectral Type & Distance & $T_{\rm eff}$ & \Rstar\/ & $\log L_{\rm bol}$ & \vinf\/ & $\Mdot_{\rm theory}$ & $N_{\rm ISM}$  \\
  & & & (pc) & (kK) & (\Rsun) & ($L_{\odot}$) & (km s$^{-1}$) & (\Msunyr) &  ($10^{22}$ cm$^{-2}$) \\
  \hline
   \zpup\ & 66811 & O4 If & 332$^{a}$ & 40.5$^b$ & 13.5$^{b}$ & $5.65^{b}$ & 2250 & $4.9 \times 10^{-6}$ & 0.01  \\
  9 Sgr & 164794 & O4 V & 1250$^c$ & 42.9$^{d}$ & 12.4$^{d}$ & 5.67$^d$ & 3100 & $2.1 \times 10^{-6}$  & 0.22  \\
  \zori\ & 37742 & O9.7 Ib & 226$^a$ & 30.5$^d$ & 22.1$^d$ & $5.58^{d}$ & 1850 & $1.2 \times 10^{-6}$  & 0.03  \\
  \eori\  & 37128 & B0 Ia & 363$^e$ & 27.5$^e$ & 32.4$^e$ & 5.73$^e$ & 1600 &  $1.2 \times 10^{-6}$ & 0.03 \\
  \xper\  & 24912 & O7.5 III & 408$^c$ & 35.0$^f$ & 14.0$^f$ & $5.42^{f}$ & 2450 &  $9.3 \times 10^{-7}$  & 0.11  \\
  \zoph\ & 149757 & O9.5 V & $135^c$ & 32.0$^f$ &  7.0$^f$ & $4.66^{f}$ & 1550 & $1.0 \times 10^{-7}$  & 0.06 \\
  \hline
\end{tabular}

{References:  $^a$\citet{vanLeeuwen2007}; $^b$\citet{Howarth2019}; $^c$\citet{GAIAEDR3};  $^d$\citet{Martins2005}; $^e$\citet{Searle2008}; $^f$\citet{Repolust2004}; all terminal velocities from \citet{Haser1995}, and all ISM column densities from \citet{Fruscione1994}. For \zoph\/ and \xper, we made adjustments to the radii from \citet{Repolust2004} based on the Gaia distances. Theoretical mass-loss rates are from the recipe in \citet{Vink2000}. 
Note that we have adopted parameters for \zpup\/ that are based on the {\it Hipparcos} distance, which leads to a smaller radius and luminosity and slightly smaller theoretical mass-loss rate than we have adopted previously (e.g.\ \citealt{Cohen10,Cohen2014b}). 
}
\label{tab:stars}
\\
\end{table*}  

There are grating spectra of about two dozen O and early B stars in the \chandra\/ archive. This includes quite a few O+O and O+WR colliding wind binary sources as well as several objects whose X-ray emission is dominated by magnetically channeled wind shock X-rays. In their comprehensive study of line profiles in \chandra\/ grating spectra of non-magnetic, non-binary O star X-ray sources \citet{Cohen2014a} analyzed twelve O stars (also including the B0 supergiant \eori) that were not known definitively to be either CWS or MCWS X-ray sources. X-ray line profile shapes are sensitive to absorption but not to emission temperature and this analysis showed that several of those stars have X-ray spectra that are indeed contaminated by CWS X-ray emission. We therefore define our sample as the subset of the \citet{Cohen2014a} sample that were determined to be dominated by EWS X-rays. We list these objects and the properties we adopt for them in Table \ref{tab:stars}. Note that we order the sample by their winds' (theoretical) mass-loss rates. The sample does not include HD 93129A, which is also primarily an EWS X-ray source, because its \chandra\/ grating spectrum is so absorbed that only a handful of lines are measurable \citep{Cohen2011} and the six-temperature DEM cannot be well constrained. 


\begin{table}
  \caption{\chandra\/ observing log}
\begin{tabular}{cccc}
  \hline
  Star & Observation ID & Exposure Time & Date \\
  & & (ks) &  \\
  \hline
   \zpup\ & 640 & 67.74 & 28 Mar 2000  \\
  9 Sgr & 5398 & 101.23 &  11 May 2005   \\
   & 6285 & 44.6 & 09 Jun 2005 \\
   & {\it total} & {\it 145.83} & \\
  \zori\ & 13460 & 142.9 & 29 Nov 2011   \\
   & 13461 & 53.0 & 10 Dec 2011 \\
   & 14373 & 46.42 & 09 Dec 2011 \\
   & 14374 & 15.34 & 06 Dec 2011 \\
   & 14375 & 36.06 & 13 Dec 2011 \\
   & {\it total} & {\it 293.72} & \\
  \eori\ & 3753 & 91.65 & 12 Dec 2003    \\
  \xper\ & 4512 & 158.8 & 22 Mar 2004     \\
  \zoph\ & 2571 & 35.41 & 04 Sep 2002    \\
   & 4367 & 48.34 & 05 Sep 2002 \\
   & {\it total} & {\it 83.75} & \\
  \hline
\end{tabular}
\label{tab:observations}
\end{table}  

The \chandra\/ High Energy Transmission Grating Spectrometer (HETGS) consists of two grating arrays -- the medium energy grating (MEG) and the high energy grating (HEG). The \heg\/ has superior spectral resolution but poorer throughput at the wavelengths where the sample stars radiate most of their X-rays and so the \meg\/ spectra are generally more useful in our analysis. We retrieved the published HETGS data available in the \chandra\/ archive in late 2017 for all the sample stars and reprocessed all the data using standard \ciao\/ (v.\ 4.9) scripts. We co-added the negative and positive first-order \meg\/ spectra, and likewise for the \heg\/ spectra. We then adaptively grouped the resultant co-added spectra, requiring at least 40 counts per bin to effectively smooth the continuum. Finally, we grouped the forbidden and intercombination lines together into one large bin for each helium-like complex as the \apec\/ X-ray emission model does not include the effects of photoexcitation from the metastable upper level of the forbidden line to the upper level of the intercombination line by the photospheric UV radiation. The model does compute the sum of the two line fluxes correctly. We present an observing log in Table \ref{tab:observations} and display the co-added, grouped \meg\/ and \heg\/ spectra, along with the best-fit models that we describe in \S4, in Figs.\/ \ref{fig:MEGdata} and \ref{fig:HEGdata}.


\section{Modeling the data} \label{sec:modeling}

\begin{figure}
\includegraphics[angle=0,width=0.45\textwidth]{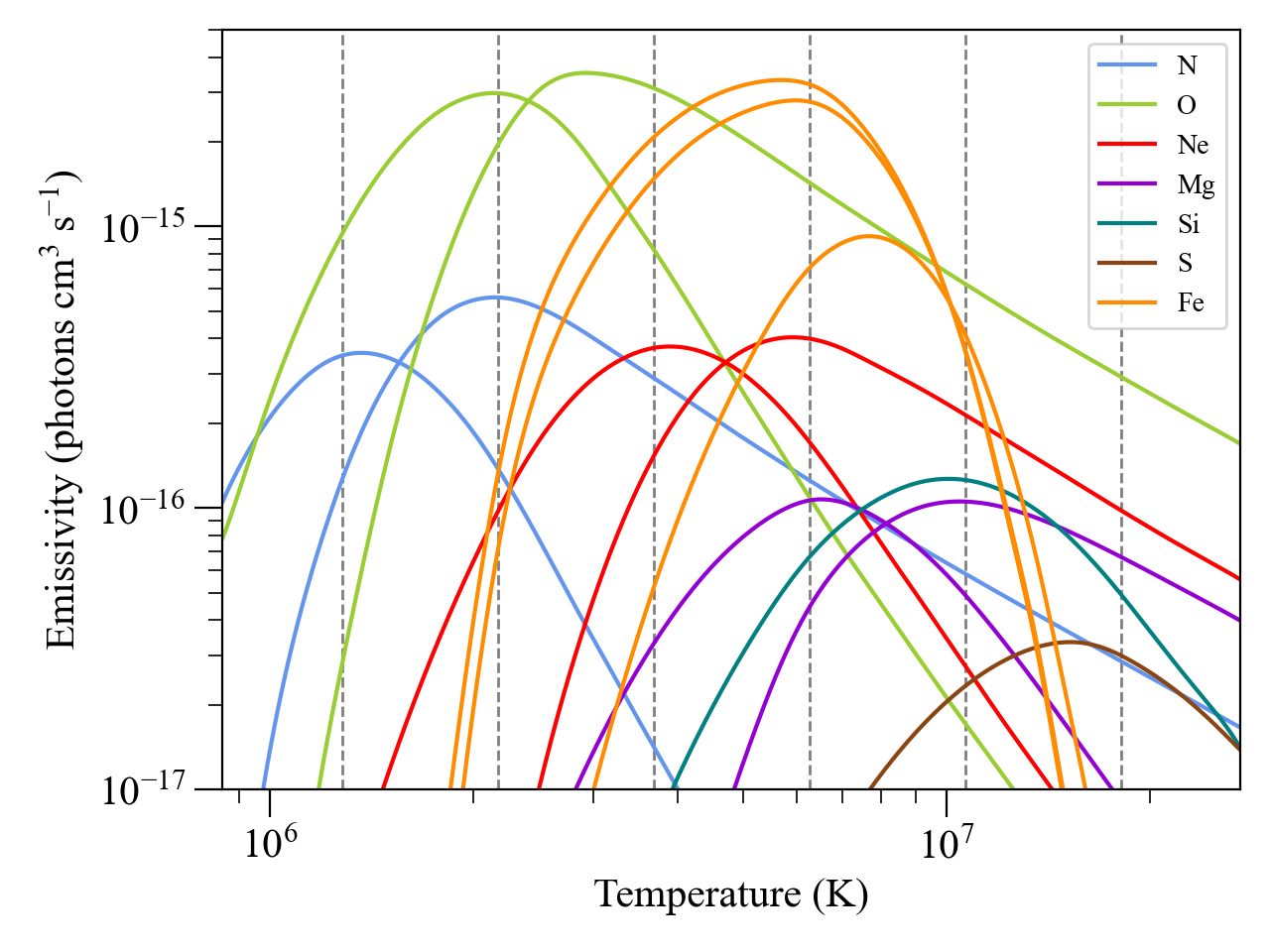}
	\caption{Emissivities of the lines seen in the sample stars' \chandra\/ spectra, taken from {\sc atomdb} \citep{Foster2012} and implemented in \bvapec, and color-coded according to species. Solar abundances are assumed here and each curve is scaled by its abundance -- multiplying any curve by the product of the electron density and hydrogen density gives the luminosity of the line. The six fixed temperatures we use in all of the model-fitting shown in this paper are indicated by the vertical dashed lines. Looking at a given vertical line and noting which spectral lines' emissivity curves it intersects gives a sense of which lines seen in the spectra are controlled by which emission model temperature component. 
	}   
	\label{fig:emissivities}
\end{figure}

We model the emission portion of our program stars' spectra with a sum of six \bvapec\/ thermal spectral emission components of fixed temperatures at evenly spaced ($0.23$ dex) logarithmic intervals from roughly $10^6$ K to $18 \times 10^6$ K ($kT = .110, .187, .318, .540, .919, 1.56$ keV). This summed model approximates the spectrum from the expected continuous temperature distribution, with the fixed temperatures separated by roughly the characteristic width of any particular line's emissivity and thus sampling the temperature distribution as finely as is reasonable. We selected the particular temperatures to sample the peaks of the emissivity functions of all important spectral lines, as shown in Fig.\ \ref{fig:emissivities}. In the formalism and implementation of \apec\/ and similar models, the emissivity has units of photons emitted cm$^{3}$ s$^{-1}$ and is multiplied by the emission measure (cm$^{-3}$) to give the luminosity as a function of wavelength. The emissivities of individual lines, like those shown in Fig.\ \ref{fig:emissivities}, are computed from a collisional-radiative non-LTE ionization and excitation model that gives the relevant level populations as a function of temperature.

The \apec\/ model \citep{Foster2012} is used for optically thin plasma in collisional ionization equilibrium, with line emission resulting from collisional excitation followed by radiative de-excitation, and continuum emission resulting from bremsstrahlung and recombination. The \bvapec\/ variant additionally accounts for Doppler broadening via convolution with a Gaussian, as well as variable abundances for up to a dozen elements. We report abundances on a linear scale with respect to the Solar standard of \citet{Asplund2009}.


We use \vwindtab, a version of \windtabs\/ \citep{Leutenegger2010} allowing variable abundances, to model the absorption from the colder portion of the wind that composes the vast majority of its mass. This absorption arises from inner-shell photoionization of few-times-ionized metals and is dominated in much of the \chandra\/ bandpass by the opacity of nitrogen and oxygen. We allow the abundances of these two elements to be free parameters of the \vwindtab\/ model and tie them to the corresponding abundance parameters in the \bvapec\/ emission model. We use the \tbabs\/ model \citep{Wilms2000} to account for interstellar absorption, using fixed hydrogen column density values taken from the literature for each star, as listed in Table \ref{tab:stars}. 

Because we fix the temperatures of all six emission components, the only free parameters of the emission model are the six normalizations, the N and O abundances, a single line width value, and a Doppler shift value. This last value is an uninteresting parameter in the context of the present work, encompassing both wavelength calibration errors and space motion of the stars. It also is affected by the line profile asymmetry caused by wind attenuation \citep{Owocki2001}. Because the N and O abundance values in the \vwindtab\/ model are tied to those in the \bvapec\/ emission model, the wind absorption model introduces only one new free parameter, the wind's fiducial mass column density $\Sigma_{\rm \ast} = \frac{\Mdot}{4{\pi}\Rstar{\vinf}}$ (g cm$^{-2}$). By measuring this quantity, we can infer the wind mass-loss rate. We note that we fix the shock onset radius and acceleration parameter of the wind velocity law at {\Ro}=1.5{\Rstar} and $\beta=1$ in the {\vwindtab} model. 

For fitting photon-counting Poissonian data the maximum likelihood estimator is the C statistic \citep{Cash1979}, and so we minimize that quantity to find the best-fit model for each dataset we analyze. Testing showed very similar results when using $\chi^2$ with Churazov weighting. Because our DEM and mass-loss rate measurements depend more on spectral lines than continua, we adaptively grouped our spectra, as described in the previous section. Despite the spectral grouping, our model fitting is dominated by systematic rather than statistical errors -- not just due to errors in the atomic and spectral models, which should be no more than about 25 per cent, cumulatively, for the prominent lines we model \citep{Foster2012,Hitomi2018}, 
but also due to the approximation of asymmetric line profile shapes as Gaussians -- and so we generally do not achieve formally good statistical fits. However, for some of the lower signal-to-noise datasets, the unweighted reduced $\chi^2$ values are close to unity, indicating formally good fits to those statistical-noise-dominated spectra. We report on parameter confidence limit estimation on that subset of our sample in the next section. 

To supplement the formal statistical model fitting, we assess the fit quality for the strongest spectral emission lines by comparing the integrated line flux of the model to that of the data. To do this we fit and subtract the continuum of the model and the data, separately, and numerically integrate the flux across the line. We then compute and display the model-to-data line flux ratios. This technique -- with results discussed in \S4.3 -- provides a good means of exploratory data analysis and also a visual representation of a model's success in reproducing the emission line fluxes in each spectrum. 


\section{Results} \label{sec:results}

\begin{figure}
\includegraphics[angle=0,width=0.47\textwidth]{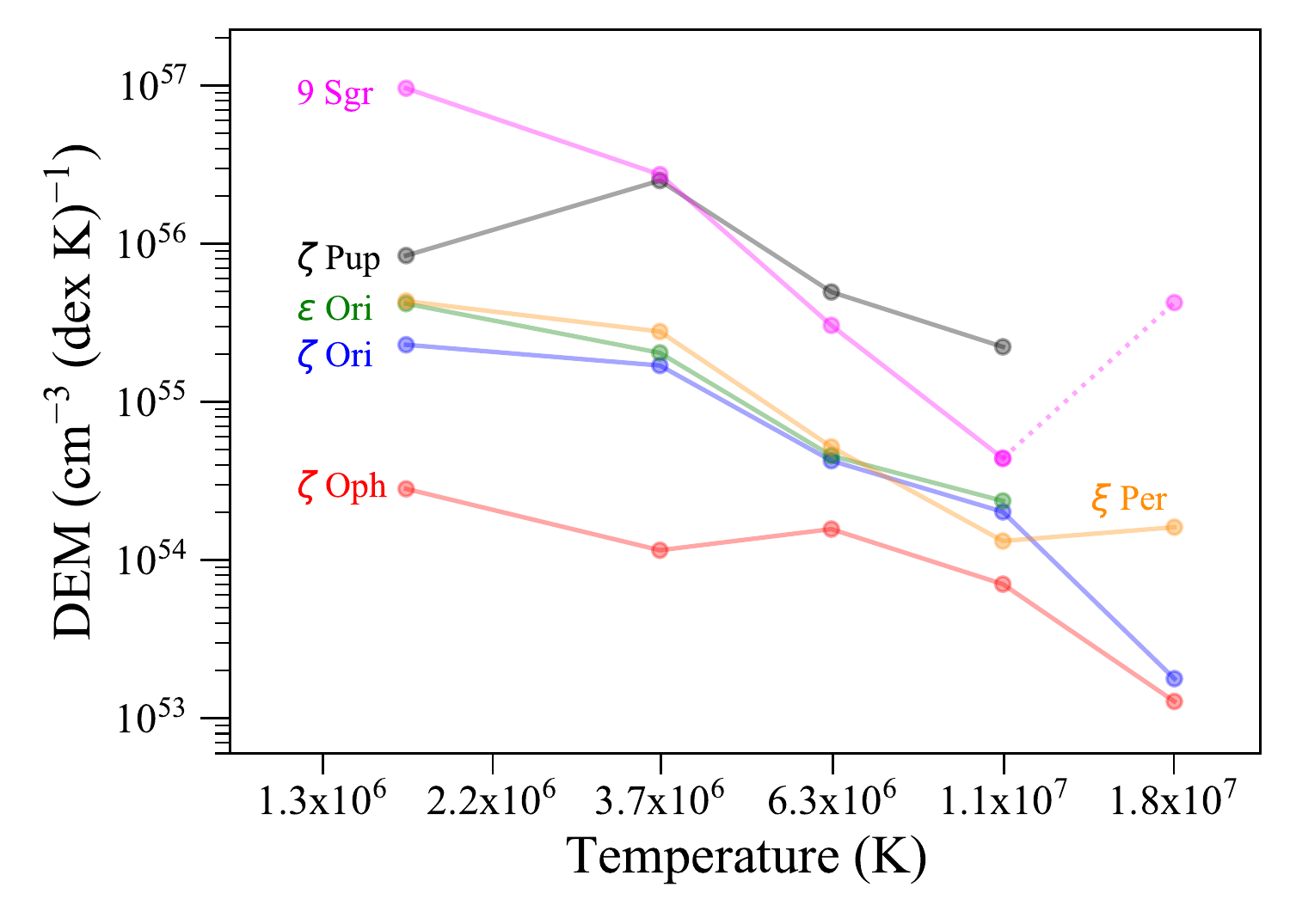}
	\caption{Emission measure distributions of the six program stars are displayed as dots representing each {\it bvapec} normalization, with line segments connecting them. For \zpup\/ and \eori\/ the emission measure in the hottest component is negligible. For \sgr\/ the hottest component has a significant contribution from CWS X-rays. For all stars we averaged and combined the two lowest temperature components into a single bin. 
	}   
	\label{fig:DEM_modified}
\end{figure}

We find similar shaped DEMs for all six program stars -- shown in Fig.\ \ref{fig:DEM_modified} -- with the emission measure a relatively smooth and strongly decreasing function of temperature such that the vast majority of the X-ray-emitting plasma has a temperature of no more than eight or nine million Kelvin. The stars' DEMs between two and 11 million K are relatively well fit by a power law, for which we find an average index of $n = -2.3$. The overall levels of the DEMs vary from star to star, with total emission measures of hot plasma ranging from $10^{54}$ cm$^{-3}$ for the star with the weakest wind (\zoph) to about $10^{56}$ cm$^{-3}$ for the stars with the strongest winds (\zpup\/ and 9 Sgr). 


\begin{table*}
  \caption{Best-fit model parameters}
\begin{tabular}{lcccccc}
  \hline
 Parameter &  \zpup\ & 9 Sgr & $\zeta$ Ori & $\epsilon$ Ori & $\xi$ Per & $\zeta$ Oph   \\
    \hline
norm$_{\rm 1}$ & 114 & $167_{-33}^{+47}$ & 0 & 0 & $21_{-17}^{+20}$ & $1.2_{-1.2}^{+21}$  \\
norm$_{\rm 2}$ & 182 & $69_{-16}^{+21}$ & 166 & 122 & $79_{-16}^{+19}$ & $58_{-16}^{+13}$  \\
norm$_{\rm 3}$ & 441 & $34_{-5}^{+4}$ & 70 & 30 & $32\pm4$ & $12_{-2}^{+5}$  \\
norm$_{\rm 4}$ & 87 & $3.7_{-1.4}^{+2.2}$ & 15 & 6.7 & $6\pm1.3$ & $17_{-3}^{+1}$  \\
norm$_{\rm 5}$ & 39 & $0.54_{-0.54}^{+0.56}$ & 8.1 & 3.5 & $1.5\pm0.5$ & $7.4_{-0.7}^{+0.8}$  \\
norm$_{\rm 6}$ & 0 & $5.2_{-0.3}^{+0.4}$ & 0 & 0 & $1.9\pm0.2$ & $1.3_{-0.4}^{+0.5}$  \\
N abundance & 17.8 & $9.5_{-1.7}^{+1.5}$ & 3.3 & 1.3 & $8.6_{-1.3}^{+1.2}$ & $2.8_{-0.5}^{+0.8}$  \\
O abundance & 0.81 & $0.71_{-0.08}^{+0.1}$ & 0.96 & 0.86 & $0.73\pm0.08$ & $0.60\pm0.1$  \\
\Sigmastar\/ & $26$  &  $10_{-1.2}^{+2}$  &  $6.9$  &  $5.9$ &  $11_{-1.7}^{+1.0}$  &  $6.2_{-0.9}^{+1.1}$ \\
$N_{\rm H}$ & 0.01 & 0.22 & 0.03 & 0.03 & 0.13 & 0.06  \\
redshift & -460 & -380$\pm40$ & -97 & -120 & -120$_{-30}^{+20}$ & -29$\pm13$  \\
line width & 790 & $1000_{-30}^{+100}$ & 680 & 600 & $880_{-20}^{+30}$ & $340_{-20}^{+10}$  \\
$N_{\rm data}$ & 593 & 190 & 772 & 247 & 227 & 229  \\
$\chi^2$ & 2208 & 114 & 3356 & 384 & 211 & 153  \\
C statistic & 2488 & 753 & 2718 & 608 & 710 & 610  \\
   \hline
\end{tabular}
\label{tab:parameters}

{The six normalizations (for fixed temperatures of $kT = .110$, .187, .318, .540, .919, 1.56 keV) are given in units of $10^{10}\cdot4\pi d^2$ cm$^{-3}$, where $d$ is the distance to the star. N and O abundances are relative to Solar \citep{Asplund2009}. The wind column mass, \Sigmastar, is in units of $10^{-3}$ g cm$^{-2}$, while the interstellar column densities are in units of $10^{22}$ H-atoms cm$^{-2}$ and are fixed at the values listed in Table \ref{tab:stars}.  The redshifts and line widths (Gaussian $\sigma$) are free parameters of the model and are given in km s$^{-1}$. The final three rows list the number of bins in the co-added and grouped MEG and HEG spectra, combined, along with the values of two fits statistics, noting that the C statistic is what is minimized in the fitting; the $\chi^2$ value is provided for illustrative purposes, to give a sense of the extent to which the fits are formally good, and likely dominated by statistical errors, versus formally bad, and likely dominated by systematics.
}
\\

\label{tab:fitparams}
\end{table*}  

All the best-fit model parameters, corresponding to the spectral models shown in Figs.\ \ref{fig:MEGdata} and \ref{fig:HEGdata} and including the normalizations corresponding to the emission measures plotted in Fig.\ \ref{fig:DEM_modified}, are listed in Table \ref{tab:parameters}. These parameters include the characteristic wind mass column density $\Sigma_*$ (g cm$^{-2}$), which is the primary adjustable parameter of the wind absorption component of the spectral model. We use these fitted $\Sigma_*$ values, along with the parameters listed in Table \ref{tab:stars}, to calculate wind mass-loss rates for each star, which we list in Table \ref{tab:mdots}.

\begin{figure*}
    \includegraphics[angle=0,width=\textwidth]{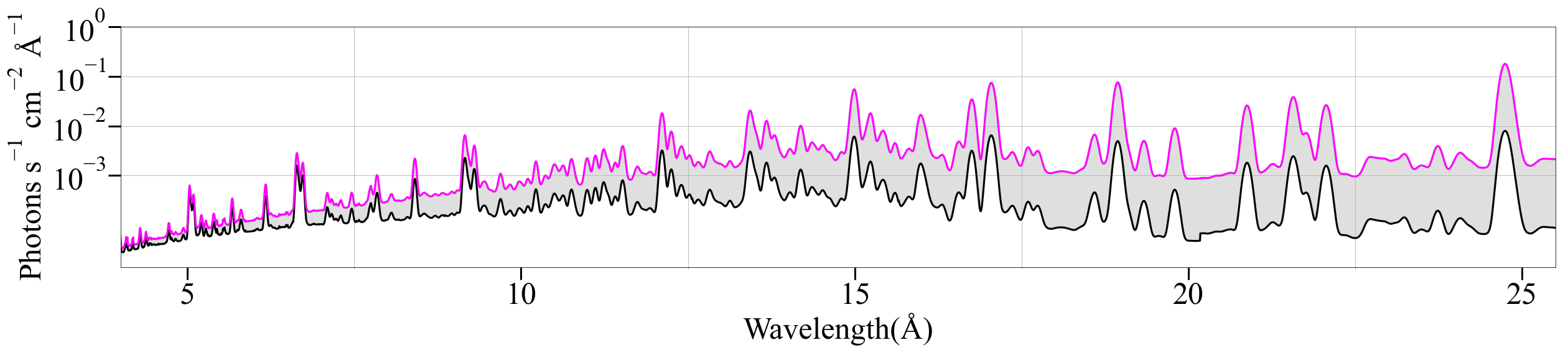}
    \caption{Best-fit model spectrum for {\zpup} (black line) compared to the intrinsic emitted spectrum, corrected for wind attenuation (pink line). The strong wavelength-dependence of the wind absorption can easily be seen, as can the very significant degree of wind absorption of the softer X-rays. Both models are corrected for ISM absorption.}
    \label{fig:Sigma0}
\end{figure*}

For all of the sample stars, wind absorption is a significant effect and for some, it dominates the appearances of the spectra. Fig.\ \ref{fig:Sigma0} compares the intrinsic (``emitted'') and emergent X-ray spectrum of \zpup, computed simply by zeroing out the $\Sigma_*$ parameter in the best-fit model. More than 95 percent of the photons emitted by the oxygen and nitrogen lines between 18 and 25 \AA\/ are absorbed by the wind.

For each star except {\sgr} we find $\log({\Lx}/{\Lbol})$ (the standard definition -- emergent \Lx, not corrected for wind absorption) to be close to the canonical value of -7, with an average of -7.1 for the other five stars. For {\sgr}, we find a value of -6.3, which, although high, is consistent with the value of -6.35 found using \xmm\/ data by Rauw et al. (2002). We calculate the X-ray fluxes by integrating the best-fit models over the 0.5 to 10 keV range, correcting for ISM attenuation, and using the distances given in Table \ref{tab:stars} to convert from flux to {\Lx}. 

To compute the X-ray luminosity generated by the shocked wind plasma, which may be more directly useful for evaluating simulations than the emergent X-ray luminosity is, we can correct for the wind attenuation by setting \Sigmastar\/ of the best-fit model to zero. The quantity calculated this way, $\log(L_{\rm X}/L_{\rm bol})_{\rm emit}$, is reported in the last column of Table \ref{tab:mdots}. The uncorrected quantity is the one typically quoted in the literature, but the corrected quantity provides information about the efficiency of the X-ray production of the wind.

In the subsections below we elaborate on the emission measures, secondary model parameters including abundances, and on the model-fitting technique itself.


\begin{table*}
\caption{Mass-loss rates and X-ray luminosities}
\begin{tabular}{ccccccc}
\hline
{Star} & Spectral & Theory: Vink et al. 2001 & Cohen et al. 2014 & This Work & { $\log ( \frac{L_{\rm X}}{L_{\rm bol}} )$ } & {$\log( \frac{L_{\rm X}}{L_{\rm bol}} )_{\rm emit}$ } \\
  & Type & ($M_{\odot}$ yr$^{-1}$) & ($M_{\odot}$ yr$^{-1}$) & ($M_{\odot}$ yr$^{-1}$) &  &  \\
\hline
$\zeta$ Pup & O4 I & $4.9\times10^{-6}$ & $1.8\times10^{-6}$ & $1.1\times10^{-6}$ & $-6.97$ & $-6.03$ \\
 9 Sgr & O4 V & $2.1\times10^{-6}$ & $3.7\times10^{-7}$ & $5.3\times10^{-7}$ & $-6.30$ & $-5.81$ \\
$\zeta$ Ori & O9.7 I & $1.2\times10^{-6}$ & $3.4\times10^{-7}$ & $3.9\times10^{-7}$ & $-7.26$ & $-6.95$ \\
$\epsilon$ Ori & B0 I & $1.2\times10^{-6}$ & $6.5\times10^{-7}$ & $4.2\times10^{-7}$ & $-7.20$ & $-6.98$ \\
$\xi$ Per & O7.5 III & $9.3\times10^{-7}$ & $2.2\times10^{-7}$ & $5.2\times10^{-7}$ & $-7.09$ & $-6.61$ \\
$\zeta$ Oph & O9 V & $1.0\times10^{-7}$ & $1.5\times10^{-9}$ & $9.3\times10^{-8}$ & $-7.13$ & $-6.92$ \\
\hline
\end{tabular}
\label{tab:mdots}

{Because we use a smaller value for the radius of \zpup\/ than was used in \citet{Cohen2014b}, the mass-loss rates in the first row are not directly comparable -- using the older value for the stellar radius, we find $\Mdot = 1.5 \times 10^{-6}$ \Msunyr\/ from the wind column density derived in this paper. That value is in agreement with the line-profile-based determination from \citet{Cohen2014b} at the 20 per cent level. The X-ray luminosities in the last column are corrected both for ISM and wind absorption whereas the values in the second-from-last column are corrected solely for ISM absorption. In both cases the X-ray luminosities are integrated over photon energies from $0.5$ to $10$ keV. 
}

\end{table*}

\subsection{The Differential Emission Measure}

Although the DEMs shown in Fig.\ \ref{fig:DEM_modified} are relatively well behaved and display important common properties -- moderately smooth distributions, consistent with steep decreasing power laws having similar slopes and very little contribution from hot plasma with temperatures much above $10^7$ K -- there are several results that bear further discussion. 

As noted in \S3, we use a six-temperature emission model, but for a number of stars, we find negligible emission in the lowest-temperature component. This can be explained by the strong degeneracy between components 1 and 2 in the effective \chandra\/ bandpass of 5-25 \AA. Fig.\ \ref{fig:emissivities} shows that only one N {\sc vi} line contributes significantly to the lowest-temperature component (component 1). This line is blended with a stronger N {\sc vii} line near 25 \AA, so these \chandra\/ grating data provide very few independent constraints on component 1 relative to component 2. For the sake of uniformity in the logarithmic temperature spacing, we include both low-temperature components in our fitting and list the results in Table \ref{tab:fitparams}, despite their relative degeneracy, but average their emission measure levels when we display the results in Fig.\ \ref{fig:DEM_modified}. 

We find a negligible normalization for the highest-temperature component in two of stars and a $norm_{6}$ at least an order of magnitude less than $norm_{5}$ in two more. 
The emission measure in the highest temperature component is, for five of the six sample stars (the exception being 9 Sgr), much less than one per cent of the total X-ray emission measure. For {\sgr} we find a somewhat larger normalization for the highest-temperature {\bvapec} component. Multiple spectral lines became weaker or disappeared when we experimented with removing this component from the spectral model and refit the data, indicating that there is some line emission from plasma with temperatures above $10^7$ K. However, the presence of this high-temperature plasma in \sgr\/ can be at least mostly accounted for by the contribution from colliding wind shock emission. The star has a binary companion with an eccentric, $P = 9.1$ yr orbit \citep{Rauw2016}. These authors conclude that the CWS contribution is modest but more manifest in the harder X-rays than in the rest of the spectrum, implying that the hottest component of the DEM we derive could be due to the CWS emission. The companion, only slightly less luminous than the primary, likely contributes about a third of the X-ray emission from its own embedded wind shocks \citep{Rauw2016}. Even with the contribution from the companion, however, the overall amount of EWS emission and the X-ray to bolometric luminosity ratio is higher for 9 Sgr than for the other stars in the sample. 


\noindent

{\zpup} also has a DEM that is modestly different from our other program stars, with a peak near 4$\times 10^6$ K rather than at lower temperatures. However, the fluxes of the three strongest lines that contribute to the two lowest temperature components (O {\sc vii}, O {\sc viii}, and N {\sc vii}) are not as well fit as most other lines, perhaps indicating that the systematic errors on the low-temperature portion of the DEM are bigger than they are elsewhere. Despite this, the DEM we derive for \zpup\/ is not very different from that of the other program stars.

\subsection{Other Model Parameters}

The N and O abundances listed in Table \ref{tab:parameters} are constrained by the N VII Ly$\alpha$ line, the O VIII Ly$\alpha$ and $\beta$ lines, and the O VII He-like line complex just shortward of 22 \AA, as well as the wind opacity contributed by N and O. While the N/O abundance ratio is super-solar for all six stars, we find only mildly non-solar abundances for {\zoph}, {\eori}, and {\zori}, each of which is known from optical and UV studies to have solar or close-to-solar abundances \citep{Cazorla2017b, Puebla2016, Blazere2015}. We found substantially super-solar N and sub-solar O in {\xper}, {\zpup}, and {\sgr}, of which {\xper} and {\zpup} are both already known to have enhanced N/O abundance ratios \citep{Martins2017, Kahn2001}, while {\sgr} is expected to have only a slightly elevated N abundance \citep{Rauw2016}. This consistency with optical- and UV-derived abundances suggests that X-ray spectral fitting can provide approximate N and O abundance determinations via the relatively simple procedure we present in this paper.

We find that the neon line fluxes are consistently among the hardest to fit, so we tried fitting models with the Ne abundance a free parameter. For most stars, we found no significant improvement to the fit, but for {\zoph} and {\eori}, allowing the Ne abundance to be free yielded significant improvements to the three measurable Ne lines and line complexes in the spectrum without worsening the fits to other lines. The best-fit values of 1.78 times solar for {\zoph} and 1.69 times solar for {\eori} indicate that these stars may have mildly enhanced Ne abundance. We note that B stars in the Solar neighbourhood have higher than Solar \citep{Asplund2009} neon abundances \citep{Alexeeva2020}. It is also possible that the errors in the atomic model in {\it apec} are larger for Ne than for other elements or that errors in neon line emissivities' temperature dependence interact with the errors in iron line emissivities. The two sets of lines form at similar temperatures but there is more contribution from iron and so it tends to dominate the fits. Excluding either the iron or the neon lines from the fitting does not significantly affect the derived DEMs. 

The global line width values listed in Table \ref{tab:parameters} correlate well with the terminal velocity of each star's wind given in Table \ref{tab:stars}. {\zoph} has an exceptionally small line width relative to its wind terminal velocity as a result of its weak wind, a phenomenon that is seen in early B stars \citep{Cohen2008,CazorlaNaze2017} and late-O main sequence stars \citep{Skinner2008,Huenemoerder2012} and is attributed to the large post-shock cooling lengths in low-density winds. The redshift values listed in Tab.\ \ref{tab:parameters} are non-zero due largely to the wind absorption effect that causes asymmetric line profiles \citep{Cohen2014a}. These values correlate well with the stars' mass-loss rates and fitted $\Sigma_{\rm \ast}$ values, as expected. 

\subsection{Fit Quality and the Model-Fitting Technique}


\begin{figure*}
	\includegraphics[angle=0,width=0.47\textwidth]{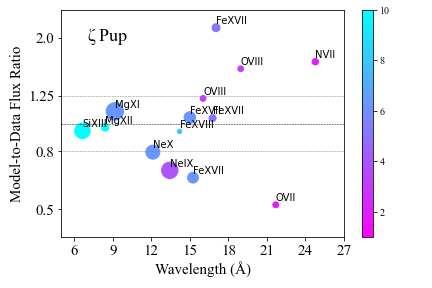} 
	\includegraphics[angle=0,width=0.47\textwidth]{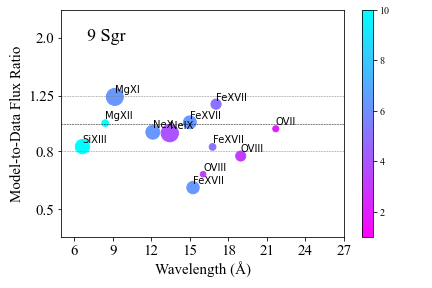}
	\includegraphics[angle=0,width=0.47\textwidth]{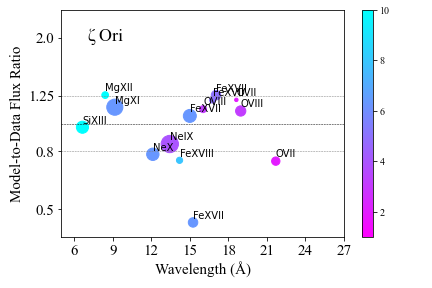}
	\includegraphics[angle=0,width=0.47\textwidth]{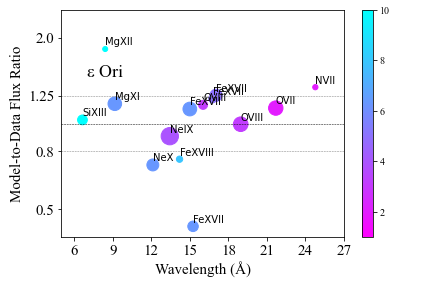}
	\includegraphics[angle=0,width=0.47\textwidth]{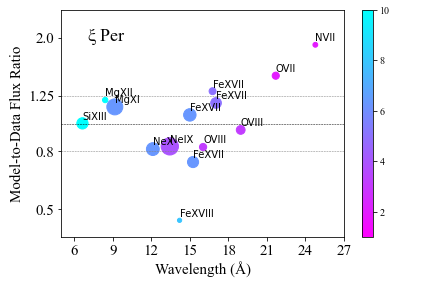}
	\includegraphics[angle=0,width=0.47\textwidth]{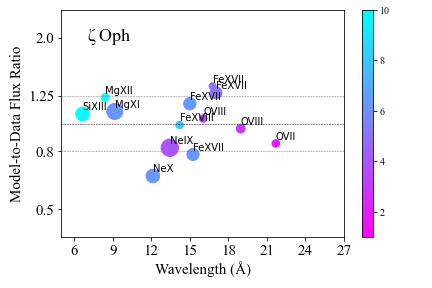}
	\caption{ Line flux ratios (model/data)  for $\zeta$ Pup, 9 Sgr, $\zeta$ Ori, $\epsilon$ Ori, $\xi$ Per, and $\zeta$ Oph (upper left to lower right). The plotted model/data ratios are a count-rate-weighted average of the MEG and HEG data (except for longer wavelength lines for which there is negligible HEG data and we include only the MEG data in the fit). The colors of the circles are proportional to the temperature of peak emissivity of the line (see color bars at the right of each panel which show the temperature in $10^6$ K), and the area of each circle is proportional to the number of counts in the line. We indicate a range of $\pm 25$ percent with the horizontal gray dotted lines about a flux ratio of unity, which would represent a perfect fit of the line flux. 
		}   
	\label{fig:dots}
\end{figure*}


\begin{figure*}
	\includegraphics[angle=0,width=0.32\textwidth]{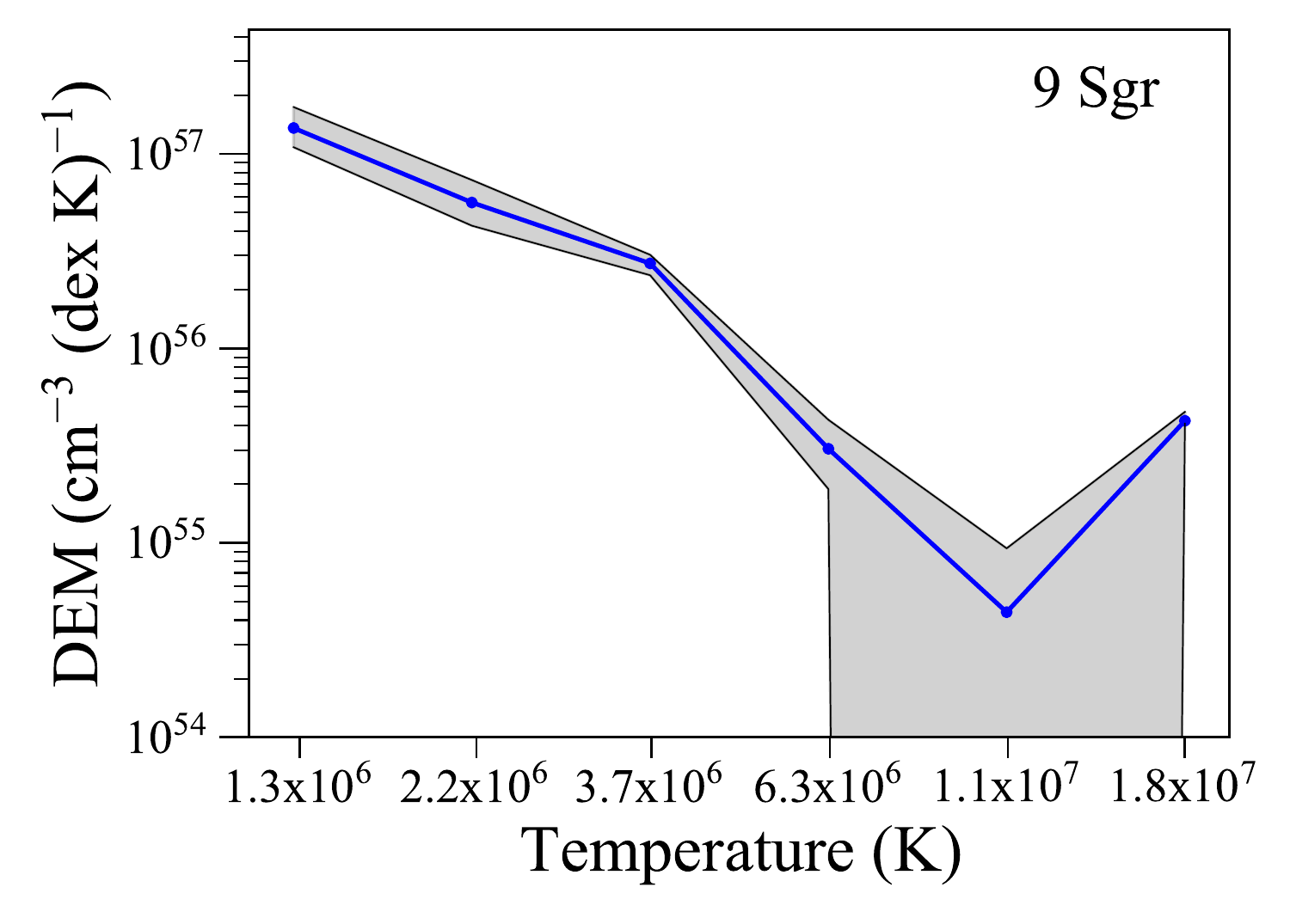}
	\includegraphics[angle=0,width=0.32\textwidth]{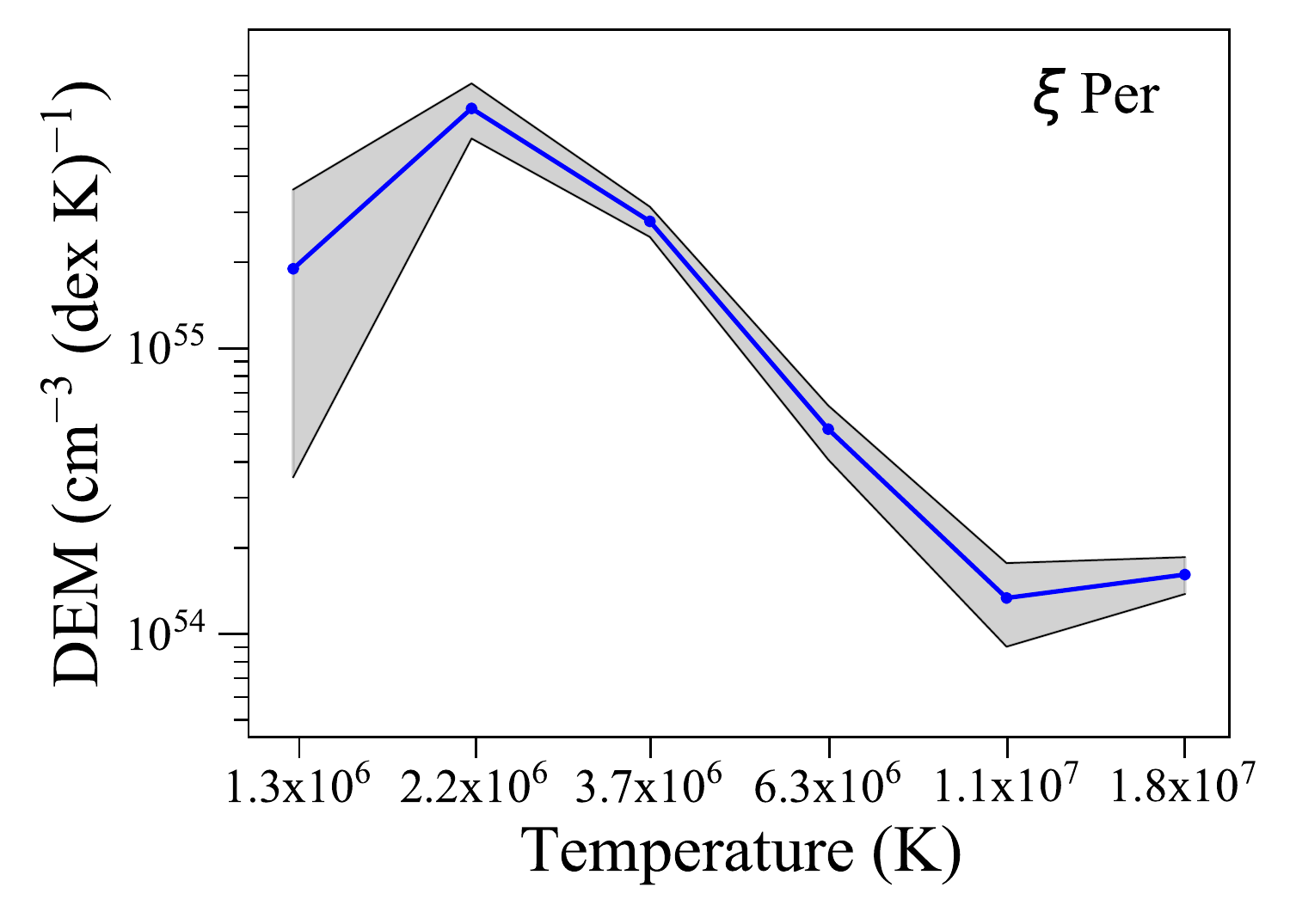}
	\includegraphics[angle=0,width=0.32\textwidth]{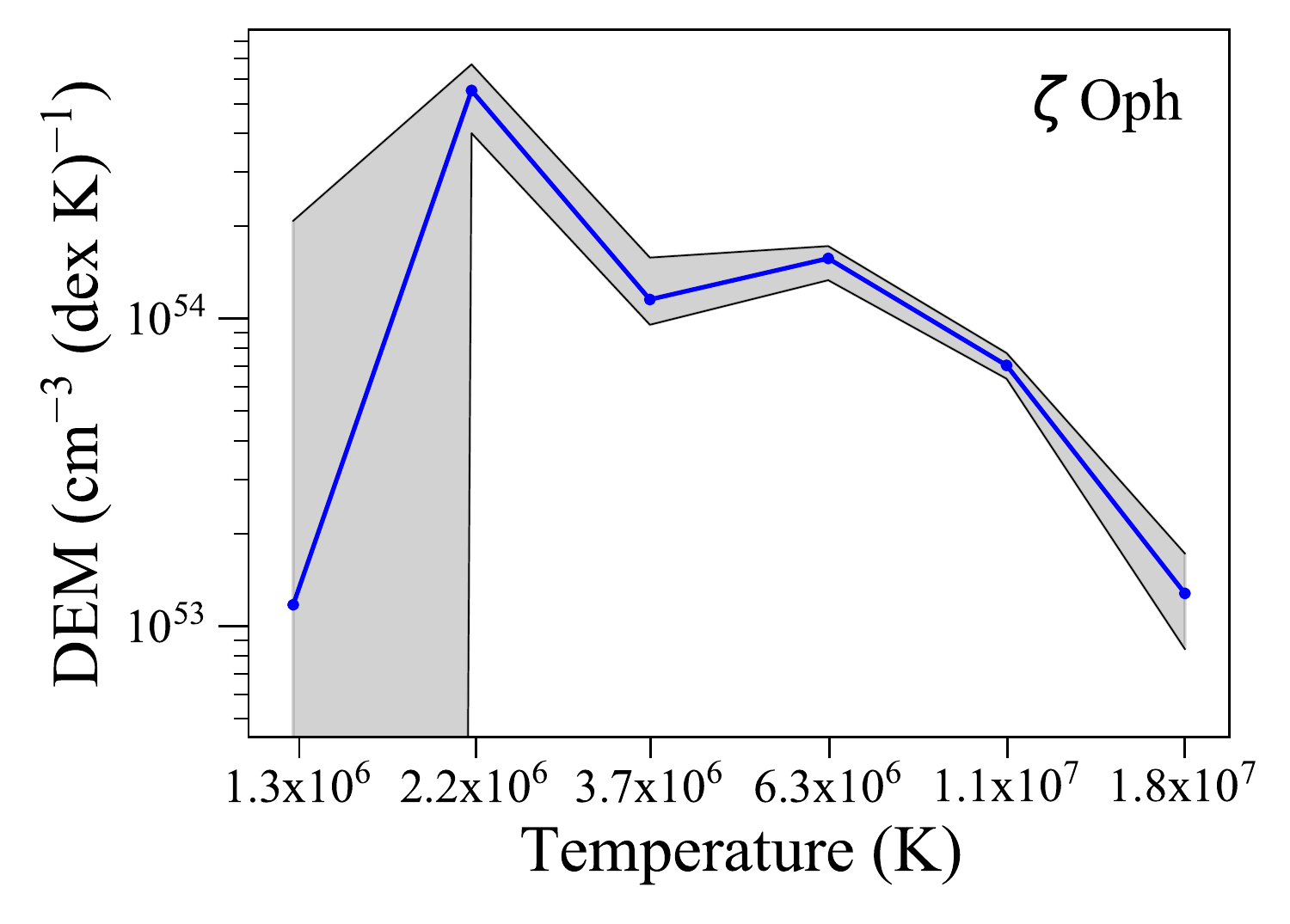}
	\caption{The 68 percent confidence limits on the DEM are shown as gray bands surrounding the best-fit DEMs (blue points and lines) for the three program stars that have statistically good fits according to the unweighted $\chi^2$ values of the best-fit models. Note that unlike in Fig. \ref{fig:DEM_modified}, we do not add together the two lowest-temperature bins but rather show them separately. 
		}   
	\label{fig:DEM_errors}
\end{figure*}

We conclude this section with a summary of some of the fit quality and technique-related results. First of all, the fitting in \xspec\/ was fast, with only slightly longer run times than more standard two-temperature {\it apec} plus ISM absorption models. The fits were not formally good -- they were dominated in each case to one extent or another by systematic errors. However, the line fluxes in all cases were well reproduced by the best-fit models. In Fig.\ \ref{fig:dots} it can be seen that nearly all lines for all six program stars had their fluxes fit to better than $\pm 25$ per cent. The figure shows that the exceptions are a small number of weak, low signal-to-noise lines and in some cases the neon lines (which led us to explore non-solar neon abundances as described above), and some of the iron lines, primarily in the spectrum of \zpup. Many of the discrepancies in the iron L-shell lines are consistent with known uncertainties in atomic models \citep{2006PhRvL..96y3201B, 2006ApJ...646..653C}. Furthermore, dielectronic satellite lines from Na-like Fe XVI are not included in \apec, but are known to be important for plasma with temperatures of a few  $\times 10^6$ K \citep{2018ApJ...864...24B}; these lines appear mostly between the 15.01 and 15.26 \AA\ lines.

To get a sense of the confidence limits on our derived model parameters -- at least those due to statistical errors -- we performed standard error analysis on the three lowest signal-to-noise spectra, for which the best-fit models give a reduced $\chi^2$ value close to unity when using standard weighting of the $\chi^2$ statistic in \xspec. We computed the 68 per cent parameter confidence limits using the $\Delta$ $C = 1$ criterion \citep{NS1989} while all the other model parameters were allowed to be free.  Fig.\ \ref{fig:DEM_errors} shows quite tight confidence limits leading to a relatively narrow band of allowed DEMs for each star. For these same three stars, we also found relatively tight confidence limits on \Sigmastar, of better than $\pm 18$ per cent of the best-fit value in each case. Presumably the statistical errors on the DEMs and wind column masses of the other three program stars are similar, although we do not repeat this exercise for those stars because their higher signal-to-noise spectra are dominated by systematic errors. 

\section{Discussion} \label{sec:discussion}
a
Our study has two main results: wind absorption of soft X-rays is a strong effect, accurate modeling of which is crucial for interpreting soft X-ray spectra of massive stars; and the intrinsic emission from our program stars appears to follow a universal differential emission measure distribution. This distribution is well approximated by a power law with a slope of about $-2.3$ that cuts off around a temperature of $10^7$ K and with an intrinsic flux that scales with the mass-loss rate of the wind. This implies that the fundamental shock physics associated with the LDI is similar for O stars with winds that span more than two orders of magnitude in mass-loss rate. The primary caveat being that the sample of stars is quite small. 

The DEM results are consistent with an earlier study of X-ray line emission in the same datasets that assessed the shock heating rate \citep{Cohen2014b} and found a power-law heating rate with a  slope of about $-3$ and marginal detection of a cut-off above $T \approx 10^7$ K. The universal DEM we present here is not as steep because the integrated radiative cooling of plasma between 1 and 10 million K is a modestly decreasing function of temperature (see e.g.\/ \citet{Foster2012}). The temperature distribution -- the DEM - is set by heating-cooling equilibrium and if both are power-law functions of temperature, the power-law DEM has an index that is the difference between the indices of the heating and cooling functions. 

Unfortunately, at present it is difficult to make quantitative comparisons between theoretical predictions of EWS properties and the DEMs we measure. Simulations that treat the radiation transport in enough detail to include the LDI are too expensive to do in more than one dimension and also include an energy equation \citep{Dessart05,Sundqvist2018}. Therefore there are no multi-dimensional simulations of LDI-induced embedded wind shocks that make detailed predictions of the shock-heated plasma DEM or the emergent X-ray spectrum. Further, there are important parameters of the simulations that seem to control the overall X-ray luminosity and the shock onset radius in the wind. These include the description of photospheric perturbations that propagate into the wind and seed instabilities \citep{Feldmeier1997} and the inclusion of limb darkening \citep{Sundqvist2013} and rotation \citep{Sundqvist2018}. It is conceivable that these physical ingredients could also affect the shock strength distribution and hence the X-ray DEM of a wind. While it is possible that stronger shocks and hotter plasma could occur in such simulations, the LDI does seem to have difficulty producing very strong shocks, and so the DEMs we find in this paper are semi-quantitatively consistent with the shock strength distributions seen in simulations \citep{Feldmeier1997,Runacres2002,Dessart05,Sundqvist2018}. Future computational advances will be required to make direct comparisons between the results we present here and numerical simulations of embedded wind shocks. 

The mass-loss rates we derive from the wind column-density parameter, \Sigmastar, are a factor of several lower than the traditional theoretical values \citep{Vink2000} and consistent with  multi-wavelength diagnostics that account for modest wind clumping \citep{Puls06}. Our broadband X-ray absorption mass-loss rates are also consistent with the X-ray line-profile-based results \citep{Cohen2014a}, with the exception of the weak-wind star \zoph\/ which shows some soft-X-ray attenuation in this study indicating a low mass-loss rate but not as low as indicated by X-ray line profile analysis \citep{Cohen2014a}. The effects of wind attenuation on the longer-wavelength portions of the sample stars' \chandra\/ spectra are generally quite significant, even for \zoph, as a comparison between the last two columns of Table \ref{tab:mdots} shows. To some extent it seems that the observed $L_{\rm x} \propto L_{\rm bol}$ trend may indeed be due to wind absorption as suggested by \citet{OC1999}, given that stars with higher mass-loss rates tend to have higher $\log(L_X/L_{bol})_{emit}$ but roughly the same $\log(L_X/L_{bol})$ as stars with lower mass-loss rates. 

\begin{figure}
	\includegraphics[angle=0,width=0.45\textwidth]{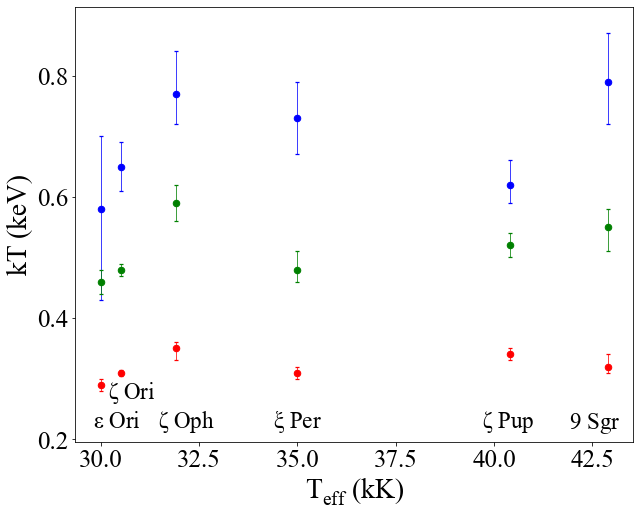}
	\caption{The line-ratio based temperatures for Si {\sc xiv/xiii}, Mg {\sc xii/xi}, and Ne {\sc x/ix} (blue, green, red) are quite constant from star to star. Note that the effective temperature scale does not map exactly onto the theoretical mass-loss rate ordering of the sample that we use throughout the paper because the Orion belt stars are supergiants and have higher mass-loss rates than \xper\/ and \zoph. 
		}   
	\label{fig:kT_vs_Teff}
\end{figure}

Fig.\/ \ref{fig:Sigma0} emphasizes not just the magnitude of the X-ray absorption effect but its strong wavelength dependence. The universal shape of the DEMs shown in Fig.\ \ref{fig:DEM_modified} suggests that the X-ray spectral hardness trend discovered by \citet{wnw2009} is largely or entirely a wind absorption effect. Following those authors, we investigate the extent of any residual  absorption-independent plasma temperature trend in the six EWS stars in our sample by fitting isothermal {\it bapec} models to just the H-like and He-like neon K$\alpha$ lines to derive a single emission temperature. This should show relatively little bias from absorption effects, since the wavelengths are similar. We repeat the exercise for Mg and Si and show the results in Fig.\ \ref{fig:kT_vs_Teff}. Our Ne results agree almost exactly with those from \citet{wnw2009}.  The X-ray ionization temperatures we find are essentially the same from star to star for each element. We tried modeling these temperatures as a linear function of photospheric effective temperature and obtained only a weak positive correlation of Ne and Mg with $T_{\mathrm {eff}}$ while finding a slightly negative correlation for Si. Notably the two stars with the strongest winds and highest X-ray luminosities -- \sgr\/ and \zpup\/ -- have X-ray ionization temperatures that are consistent with those of the other stars. This further indicates that there is little or no intrinsic X-ray emission temperature trend with stellar spectral type or effective temperature or wind mass-loss rate -- although a sample of six stars is admittedly quite small.   


Perhaps it is not surprising that the temperature distribution of EWS plasma does not scale with the stellar effective temperature. Wind properties scale primarily with the luminosity of the star and shock properties seem much more likely to be affected by the nature of the perturbations at the base of the wind than by small differences in the stellar effective temperature \citep{Feldmeier1997}. By their nature, the reverse shocks formed by the LDI have a strength governed by how anomalously fast the pre-shock wind flow is accelerated above the local ambient velocity. 

The X-ray spectral analysis technique we present here is simple to use and though in some ways is likely to be less accurate than the results of customized multi-wavelength non-LTE modeling of an individual star's wind (e.g.\/ \citealt{Herve2013}), it contains more of the first-order physics than do generic emission and absorption spectral models. The line flux model-to-data ratio diagnostic (Fig.\ \ref{fig:dots}) is useful both for exploratory analysis and to avoid over-interpretation of data. The particular six temperatures we use, with their good sampling of line emissivity functions and spacing of 0.23 dex in temperature produce relatively smooth DEMs, which are physically plausible given that shock-heated plasma radiates at all temperatures below the initial shock temperature over the time it takes to cool back down to the ambient wind temperature.  

In the future, when X-ray spectral missions observe numerous O stars, the model-fitting procedure we have presented here can be used to easily and consistently extract physically meaningful parameters from high-quality data sets and likely estimate mass-loss-rates and temperature distributions even from lower-quality datasets. It also holds out the promise of providing a complementary means for determining nitrogen and oxygen abundances in O stars, especially as applied to \xmm\/ \rgs\/ spectra, which, for low ISM column density sources, includes a measurement of both important ionization stages of nitrogen, while the \chandra\/ HETGS includes only one.

\section{Conclusions}

The technique we present here -- approximating the continuous DEM with six specific fixed temperatures and using a variable abundance, Gaussian-broadened thermal spectral emission model in combination with the {\it vwindtab} X-ray transport model for the wind attenuation -- is a fast and relatively easy way to derive physically meaningful wind model parameters within a standard fitting package like \xspec. Applying this technique to six O stars (including one B0 supergiant) yields several interesting quantitative results, centering on the strong effect of soft X-ray wind absorption. 

(i) We show that for all six stars, more than one-third of the X-rays generated in the \chandra\/ HETGS bandpass are absorbed by the wind, and much more than that for the more luminous stars with the higher mass-loss rates. Modeling this strongly wavelength-dependent attenuation is critical for extracting parameters of the shock-heated X-ray emitting component of the wind plasma. 

(ii) The emission parameters yield differential emission measures for the program stars that are smooth and strongly decreasing functions of temperature, having almost no plasma with temperatures above $10^7$ K. The emission measure levels are consistent with a scaling proportional to the stars' mass-loss rates (\sgr\/ lies somewhat above this relationship), as is expected for radiative shocks. The shapes of all the star's DEMs are quite similar, suggesting a universal nature to embedded wind shock physics in O stars across more than an order of magnitude in wind mass-loss rate.

(iii) In addition to measuring the DEMs, we measure the wind mass-loss rates from the attenuation signatures in the spectra and find values consistent with other recent determinations, including X-ray line profile based measurements, providing yet more confirmation that O star wind mass-loss rates are lower than theoretical predictions. 

(iv) We also measure nitrogen and oxygen abundances, finding elevated N/O ratios in several of the stars. And we measure a characteristic wind velocity for each star via their X-ray line widths.

The technique we have presented here will be useful when future high-resolution X-ray spectral missions produce large numbers of even intermediate-quality spectra of O stars, as it will enable the straightforward measurements of DEMs, mass-loss rates, elemental abundances, and wind speeds. 

\section*{Acknowledgments}

The scientific results in this article are based on data retrieved from the \chandra\/ data archive. Support for this work was provided by the National Aeronautics and Space Administration through grant AR2-13001A and TM3-14001B to Swarthmore College. VVP, GD, and JW were also supported by the Physics and Astronomy Department and the Provost's Office of Swarthmore College via the Vandervelde-Cheung and Eugene M.\ Lang Summer Research Fellowships. MAL acknowledges support from NASA's Astrophysics Program. We also thank the anonymous referee for their helpful suggestions. 

\section*{Data Availability}

The X-ray spectral data underlying this article are available in the Chandra Data Archive at \url{https://cxc.cfa.harvard.edu/cda/}, and are uniquely identified with the observation identifiers (Obs IDs) listed in Table \ref{tab:observations}. 


\bibliographystyle{mnras}
\bibliography{apecwindtabs}

\end{document}